\documentstyle[aps,twocolumn]{revtex}
\begin{document}

\title{What happens to spin during the $SO(3)\to SE(2)$ contraction?\\
(On spin and extended structures in quantum mechanics)}
\author{Marek Czachor~\cite{*}}
\address{
Research Laboratory of Electronics
\\
Massachusetts Institute of Technology
\\
Cambridge, MA 02139
}
\author{Andrzej Posiewnik}
\address{
Instytut Fizyki Teoretycznej i Astrofizyki\\
Uniwersytet Gda\'nski\\
ul. Wita Stwosza 57, 80-952 Gda\'nsk, Poland
}
\maketitle
\begin{abstract}
As is known, a Lie algebra of a little
group of a timelike four vector is not
equal to $so(3)$ unless spacelike components of the vector vanish.
In spite of this fact the algebra can still be interpreted
as the angular momentum algebra, as can be shown with the explicit example
of the Dirac equation. The angular momentum corresponds to the even part
of the Dirac spin operator. Its eigenvalues in directions perpendicular
to momentum decay to zero in the infinite momentum/massless limit. This
explains physically why  only extremal helicities survive the
massless limit. The effect can be treated as a result of a
Lorentz contraction of an extended particle. A natural measure of
this extension is introduced for massless particles of any spin.
It is shown that such particles can be interpreted as circular strings
whose classical limit is described by Robinson congruence.
Finally, as an application of the even spin, we formulate the
Einstein-Podolsky-Rosen-Bohm {\it Gedankenexperiment\/} for
Dirac electrons.
\end{abstract}

\section{INTRODUCTION}

It is known that the transition from the rotation group $SO(3)$ to the
group $SE(2)$ of two-dimensional Euclidean motions can be obtained as
a continuous In\"onu-Wigner contraction of
 respective Lie algebras \cite{3,ABF,KN}.
This contraction can be explained in physical terms simply as an abstract
counterpart of the relativistic Lorentz contraction of a moving body and
it can be also shown that the same effect is obtained both in the massless
($m\to 0$) and infinite momentum ($|{\vec p}|\to \infty$) limits.
This result is intuitively quite natural since a light
cone is the large-momentum asymptotics of any $m>0$ mass hyperboloid.
Still, even though the
$SO(3)\to SE(2)$ transition is now well understood from a
 mathematical point of view, it seems that no definite physical
interpretation of this fact is generally accepted.

We know that the Lie algebra $so(3)$ corresponds physically to an
angular momentum. On the other hand there exist at least
two physical interpretations
of the translation subalgebra of $e(2)$. First, the generators of
translations can be naturally associated with the two-component position
operator for massless fields \cite{ABF} localizing massless particles
in a plane perpendicular to their momentum. Second, if one considers
the action of a four-dimensional nonunitary representation of
$SE(2)$ on the electromagnetic four-potential, it can be
shown that the ``translation'' operators generate gauge transformations
\cite{KN,KN1}. Such results are rather confusing as there is no clear
explanation for the continuous deformation of the
angular momentum into position or a generator
of gauge transformations.
Of course, it is possible that for massless fields
 various physically different
observables may satisfy the $e(2)$ algebra since massless unitary
representations of the Poincar\'e group are generated by a one-dimensional
representations of $E(2)$ so that
not much room is left for different possible algebraic
structures.

We would like to show in this paper that it is possible to interpret
the contraction $so(3)\to e(2)$ also as the continuous
deformation of the angular momentum of a relativistic particle.
The Lie algebra corresponding to the intermediate momentum ${\vec p}$,
with $0<|{\vec p}|<\infty$, is neither $so(3)$ nor $e(2)$ but, as we
shall see, there is a natural way of treating it as the relativistically
generalized angular momentum algebra.

To make our analysis as explicit as possible we shall
discus the mentioned limits in the context of the Dirac equation.
It will be shown that  relativistic spin  can be naturally
represented by an ``even'' spin operator which reduces to
ordinary nonrelativistic spin for a  particle at rest and whose
components in directions perpendicular to momentum decay to
zero as the momentum increases.
In the infinite-momentum/massless limit the operator
``points'' in the momentum direction.
The three components of the new spin
commute with the free Hamiltonian so that one can consider projections of
spin in any direction even for a particle moving with some
well defined and nonvanishing momentum. This property will be used
for deriving a relativistic version of the Einstein-Podolsky-Rosen-Bohm
paradox for two spin-1/2 particles. We shall also discuss the relationship
between the even spin operator and the Pauli-Lubanski vector, showing
that the infinite-momentum and massless limits are equivalent for the
first one but not for the latter.

We begin our analysis with  a brief summary of properties of
little algebras corresponding to any timelike or null four-momentum.

\section{LITTLE ALGEBRAS FOR ANY $p$ WITH
 $p^\alpha p_\alpha \ge 0$}
\label{Sec.1}

 A Lorentz  transformation
\begin{equation}
\Lambda(\vec \mu,\vec \nu)=e^{-i\vec \mu\vec J -i\vec
\nu \vec K} \,{\buildrel \rm def\over =}\,
e^{-iL(\vec \mu,\vec \nu),}\label{1}
\end{equation}
where $\vec J$ and $\vec K$ are, respectively, the
generators of rotations and boosts, leaves a four-vector $p$ unchanged if
\begin{equation}
L(\vec \mu,\vec \nu)p=i(\vec \nu\cdot\vec p, \,\vec \nu
p_0 +\vec \mu\times\vec p)=0.\label{2}
\end{equation}
If $p_0=0$ then either $p$ is massless and $p=0$ or $p$ is
spacelike. Both cases can be regarded as physically meaningless.
So let us assume that $p_0\ne 0$. (\ref{2}) is then satisfied if
\begin{equation}
\vec \nu p_0 =-\vec \mu\times\vec p.\label{3}
\end{equation}
Substituting (\ref{3}) into (\ref{1}) we find that the little group is
three-parameter and generated by
\begin{equation}
\vec L=\vec J- {\vec p\over p_0}\times \vec K.\label{4}
\end{equation}
(\ref{4}) satisfies the algebra
\begin{equation}
\vec L\times\vec L=i\vec J - i{\vec p\over p^2_0}(\vec p\cdot
\vec K).\label{5}
\end{equation}
Let now $\vec n=\vec p/|\vec p|$, $\vec m$ be orthogonal
to $\vec n$, $|\vec m|=1$, and $\vec l=\vec n\times\vec m$ and let
 $A_1=\vec A\cdot\vec m$, $A_2=\vec A\cdot\vec l$ and $A_3=\vec
A\cdot\vec n$ for any three-component vector $\vec A$.
We get
\begin{eqnarray}
L_1 &=& J_1 + K_2{|\vec p|\over p_0} \nonumber\\
L_2 &=& J_2 - K_1{|\vec p|\over p_0}\label{6}\\
L_3 &=& J_3 \nonumber
\end{eqnarray}
and (\ref{5}) implies
\begin{eqnarray}
{[ L_1, L_2 ]} &=& i{p^\alpha p_\alpha\over p^2_0} L_3 \nonumber\\
{ [ L_3, L_1 ]} &=& i L_2 \label{7}\\
{ [ L_2, L_3 ]} &=& iL_1. \nonumber
\end{eqnarray}
Eqs. (\ref{6}) and (\ref{7}) mean that the Lie algebra of the little group
is parametrized by $p$. In the massive case we, in general, do
not obtain $so(3)$; in fact this is the case only if $|\vec
p|=0$, that is when we consider the little group of a rest frame
four-momentum. Such a rest frame always exists for $m>0$. For
$m=0$ we have
\begin{equation}
p_0=\pm |\vec p|.\label{8}
\end{equation}
and (\ref{6}) is indeed the Lie
algebra $e(2)$. Note also that ${|\vec p|\over p_0}\to \pm 1$ in the
infinite momentum limit and, as expected, the algebra is again
$e(2)$. The same conclusions follow from the $m\to 0$ and $|\vec
p| \to\infty$ contractions of (\ref{7}). The two contractions are
equivalent and the contracted algebra is $e(2)$. A reader interested
in a geometrical interpretation of these contractions is refered to
the book by Kim and Noz (p. 200 in \cite{KN}).

The form (6) of the generators leads to a difficulty in a direct
physical interpretation, as $L_1$ and $L_2$ are not
Hermitian for finite dimensional representations of $SL(2,C)$
and nonvanishing $\vec p$. For photons represented by a four-potential
these operators generate gauge transformations.
 For the rest frame four-momentum of massive particles such
representations correspond to spin. However, what is a physical
meaning of this algebra for particles that are not at rest? In
order to answer this question we must first understand in what
way the spin operator enters relativistic  quantum mechanics.

\section{SPIN OF THE ELECTRON AND THE DIRAC EQUATION}
\label{Sec.2}

The  Uhlenbeck and Goudsmith idea of an internal angular
momentum of the electron was formally introduced to quantum
mechanics by Pauli in 1927 \cite{7}. Pauli added a new
interaction term to  the Schr\"odinger equation in order to
explain a behavior of electrons in a magnetic field. This
approach was successful but, in fact, no other
justification of the concept of spin existed at that time. A
year later Dirac formulated his relativistic wave equation
\cite{8}. He assumed that the equation
(1)  has a Schr\"odinger form $i\partial_t\Psi=H\Psi$ where
$H$ does not contain time derivatives,
(2)  factorizes the Klein-Gordon equation,
(3) is relativistically covariant,
and found that no single-component wave function can satisfy
such requirements. The additional degrees of freedom present in
the multicomponent wave function could be interpreted physically by
a non-relativistic approximation where positive energy solutions
of the Dirac equation were shown to satisfy
 the equation postulated by Pauli. The Pauli spin operator was
found to be an ``internal" part of the generator of rotations
restricted to  ``large" components of a bispinor.

In this way the  spin operator was identified, from a
mathematical viewpoint, with the spinor part of the
generator of rotations.

The first difficulty met in relativistic interpretation of this
operator was the fact that, contrary to the nonrelativistic
case, the components of spin were not constants of motion even
for a free particle (unless in a rest frame).
 In the Heisenberg picture the spin operator
of the free electron satisfies the following precession equation
\begin{equation}
\dot {\vec S}=\vec \omega\times\vec S, \label{9}
\end{equation}
where $\vec \omega=-2\gamma^5\vec p$ and only the projection of
$\vec S$ on the ``precession axis" $\vec p$, the helicity, is
conserved.
The total angular momentum $\vec J$ also commutes with the Dirac
Hamiltonian and the purely spin part of $\vec J$ can be
extracted by $\vec n\cdot\vec J=\vec n\cdot\vec S$, where
 $\vec n=\vec p/|\vec p|$. The
existence of the conserved helicity is sufficient for
representation classification purposes and  physical
applications in, for instance, the Clebsh-Gordan coefficients problem.
The components of spin in directions other than $\vec n$ are
rarely needed. In the last section of this paper we shall
consider one such case, namely the Einstein-Podolsky-Rosen-Bohm
{\it Gedankenexperiment\/} and the Bell inequality for
relativistic electrons.

The general theory of representations of the Poincar\'e group
shows that the group possesses two Casimir operators: the mass
$P^\alpha P_\alpha$ and the square $W^\alpha W_\alpha$ of the
Pauli-Lubanski vector
\[
W_\alpha={1\over 2}\epsilon_{\alpha\beta\gamma\delta}M^{\beta\gamma}
P^\delta
\]
where $M^{\beta\gamma}$ are generators of $SL(2,C)$. $W_\alpha$
commutes with $P^\beta$ for all $\beta$ hence, in particular, with
$P^0$. The Pauli-Lubanski vector appears naturally in the theory
because it, in fact, generalizes the generators of the little
group described in Sec.~\ref{Sec.1}. The Casimir $W^\alpha W_\alpha$
has eigenvalues $m^2 j(j+1)$ where $m$ and $j$ are, respectively,
 the rest mass and
the modulus of  helicity of the irreducible representation in question.

In standard approaches to relativistic field theories it is
often stated that $W_\alpha$ {\it is\/} the covariant
generalization of spin. Its square is defined as the following
element of the enveloping field of the Poincar\'e Lie algebra
\[
{W^\alpha W_\alpha\over P^\beta P_\beta}.
\]
We have, therefore, two possibilities of introducing the spin
operator in Poincar\'e invariant theories. The Pauli-Lubanski
vector has the advantage of having four conserved components and
is closely related to the generators of the little group of a
momentum four-vector. The dimension of $W_\alpha$ is, however,
energy times
angular momentum so its relationship to Pauli's 1927 spin is
not evident.

\section{THE PAULI-LUBANSKI VECTOR
 AND ELECTRON'S SPIN}
\label{Sec.3}

In the following discussion we will work in the momentum
representation and units are chosen in such a way that $c=1=\hbar$.
 The bispinor parts of generators are $S^{\alpha\beta}=
{i\over 4}[\gamma^\alpha,\,\gamma^\beta]$ and the generators of
$SL(2,C)$ are
\[
M^{\alpha\beta}=x^\alpha p^\beta -x^\beta p^\alpha +S^{\alpha\beta}.
\]
Let $\vec S$, $\vec \alpha$, $\vec J$ and $\vec K$ be defined by
\begin{eqnarray}
S^{kl} &=&\epsilon^{klm}S^m\nonumber\\
S^{0k} &=&{i\over 2}\alpha^k\nonumber\\
   J^m &=&\epsilon^{mkl}M^{kl}\nonumber\\
   K^m &=&M^{0m}\nonumber
\end{eqnarray}
where $\epsilon^{klm}$ is the three dimensional Levi-Civita
symbol. The explicit form of the generators of the Poincar\'e group
is
\begin{eqnarray}
P^0 &=& H=\vec \alpha\cdot\vec p + m\gamma^0\nonumber\\
\vec P &=& \vec p\nonumber\\
\vec J &=&\vec x\times\vec p +\vec S\nonumber\\
\vec K &=&t\vec p - \vec x H +{i\over 2}\vec \alpha \nonumber
\end{eqnarray}
The Pauli-Lubanski vector is
\begin{eqnarray}
W^0 &=& \vec J\cdot\vec p\nonumber\\
\vec W &=&{1\over 2}(\vec S H +H \vec S).\label{10}
\end{eqnarray}
In the rest frame we find $W^0=0$ and $\vec W=m \vec S
\gamma^0$. $\vec W$ is therefore spin times the
rest mass operator. In the subspace of the ``large",  positive
energy two component spinors it is indeed proportional to
Pauli's spin.

In order to understand the physical meaning of the
Pauli-Lubanski vector let us multiply (\ref{10}) from the right by
$H^{-1}$ (this operator is well defined for massive fields; for massless
fields it exists in the subspace of nonzero momenta). We get
\begin{eqnarray}
\vec {S}_p &=&\vec W H^{-1}= {1\over 2}(\vec S +\lambda\vec S\lambda)
\label{11}\\
&=&\Pi_+\vec S\Pi_+ +\Pi_-\vec S\Pi_-\nonumber
\end{eqnarray}
where $\lambda$ is the sign of energy operator and $\Pi_\pm=
{1\over 2}(1\pm\lambda)$ project on positive $(+)$ or negative $(-)$ energy
solutions. The operator $\vec {S}_p$ is therefore the
so-called even part of the generator of rotations $\vec S$. The
decomposition of operators into even and odd parts is well
known
in first
quantized approaches to the Dirac equation \cite{9,DE}.
  The even parts of operators are effectively the
parts that contribute to average values of observables
calculated in states of a definite sign of energy.
The even spin operator occurs naturally in the context of
{\it Zitterbewegung\/} and the magnetic-moment operator
of the Dirac electron \cite{b4,b5}.

All these facts suggest that the even spin operator, which is
somehow in between the two notions discussed above, might be the
correct candidate for the electron's spin. All the three
components of $\vec {S}_p$ commute with $H$ so a projection
of $\vec {S}_p$ in any direction is a constant of motion.

The explicit form of $\vec {S}_p$ for free  electrons moving
with momentum $\vec p$ is the following
\begin{equation}
\vec {S}_p={m^2\over p_0^2}\vec S +{|\vec
p|^2\over p_0^2} (\vec n\cdot\vec S)\vec n +{i m\over 2p_0^2}
\vec p\times\vec \gamma \label{12}
\end{equation}
where $m^2=p^\alpha p_\alpha$. Its components
\begin{eqnarray}
\vec m\cdot\vec {S}_p &=&{m^2\over p_0^2}\vec m\cdot\vec S
-{i m |\vec p|\over 2p_0^2}
\vec l\times\vec \gamma={S}_{p1}\nonumber \\
\vec l\cdot\vec {S}_p &=&{m^2\over p_0^2}\vec l\cdot\vec S
+{i m |\vec p|\over 2p_0^2}
\vec m\times\vec \gamma={S}_{p2}\label{13}\\
\vec n\cdot\vec {S}_p &=&\vec n\cdot\vec S={S}_{p3} \nonumber
\end{eqnarray}
satisfy the Lie algebra (\ref{7})
\begin{eqnarray}
{[ {S}_{p1},\, {S}_{p2} ]}
&=& i{m^2\over p_0^2}{S}_{p3}\nonumber\\
{[ {S}_{p3},\, {S}_{p1} ]}
&=&i{S}_{p2}\label{14}\\
{[ {S}_{p2},\,  {S}_{p3} ]}
&=&i{S}_{p1}.\nonumber
\end{eqnarray}
It follows that the even spin operator (\ref{13}) is a Hermitian
representation of the algebra (\ref{7}) although a direct substitution
of generators of $({1\over 2},0)\oplus(0,{1\over 2})$ to (\ref{6})
would not lead
to Hermitian matrices. The eigenvalues of $\vec a\cdot\vec {S}_p$,
for any unit $\vec a$, are
\begin{equation}
s_{\vec a}= \pm{1\over 2}{\sqrt {(\vec p\cdot\vec a)^2 + m^2}\over |p_0|}
\label{15}
\end{equation}
and the corresponding eigenvector  in a standard
representation is
\[
\Psi_\pm^{\vec a}= N
\left(\begin{array}{c}
\sqrt {|p_0| + m}\bigl( (|s_{\vec a}|+{1\over 2}\vec a\cdot\vec
n)w_\pm \pm {m \vec a\cdot\vec n\over 2|p_0|}w\mp\bigr) \\
\sqrt {|p_0| - m}\bigl(\pm (|s_{\vec a}|+{1\over 2}\vec a\cdot\vec
n)w_\pm - {m \vec a\cdot\vec n\over 2|p_0|}w\mp\bigr)
\end{array}\right)
\]
where $w_\pm$ satisfies $\vec n\cdot\vec \sigma w_\pm=\pm w_\pm$.

In the rest frame the eigenvalues $s_{\vec a}$ are $\pm{1\over 2}$ for
any $\vec a$. $s_{\vec a}$
tend to 0 for both $m\to 0$ and $|\vec
p|\to\infty$, if  $\vec a\cdot\vec p=0$.

The transition from $|\vec p|=0$ to $|\vec p|=\infty$
deforms continuously $su(2)$ into $e(2)$ and the spin operator
$\vec {S}_p$ becomes parallel to the momentum direction. The
latter phenomenon can be deduced from either  (\ref{12}) and (\ref{13})
 or the
discussed limits of (\ref{15}).

The above limits must be understood in terms of Lie algebra
contractions. Physically the infinite momentum limit is more
reasonable than $m\to 0$. It means that the greater
velocity of a particle, the less ``fuzzy" are the components of spin
in directions perpendicular to momentum. Intuitively, the
particle  becomes
flattened by the Lorentz contraction
 so that
 contributions to the intrinsic angular momentum from rotations
around directions perpendicular to $\vec p$ become  smaller
the greater is the flattenning.

For $m$ equals exactly zero, two of the three components of spin
vanish which agrees with the fact that the only self-adjoint
finite dimensional representations of $e(2)$ are one
dimensional. Physically this effect can be again explained by the
Lorentz contraction: A massless particle is completely
flattenned and its ``intrinsic" angular momentum can
result only from rotations
in the plane perpendicular to $\vec p$.

Equations (\ref{11}) and (\ref{15}) imply that the
eigenvalues of $\vec a\cdot \vec W$  are
\[
w_{\vec a}=p_0\,
s_{\vec a}=\pm{1\over 2} {p_0\over |p_0|}\sqrt {(\vec p\cdot\vec a)^2 + m^2}
\]
and the eigenvectors are identical to those of $\vec a\cdot \vec
{S}_p$.
 For $\vec W$ the massless and infinite momentum
limits are not equivalent. Indeed, let $\vec a\cdot\vec p=0$.
Then $w_{\vec a}=\pm{1\over 2} m\neq 0$ for any $\vec p$ and
$w_{\vec a}=0$ for $m=0$. It follows that the Pauli-Lubanski
vector, as opposed to $\vec
{S}_p$, cannot be used for a unified treatment
of spin in both massive and massless cases. The same concerns
the seemingly natural and covariant choice of $(W_0/m,\vec
W/m)$ as the relativistic spin four-vector. This property of
$\vec W$ explains the following apparent paradox. The
``polarization density matrix" (normalized by ${\rm Tr}\rho=2m$)
 for the Dirac ultra-relativistic
electron can be written as \cite{10}
\[
\rho={1\over 2} p^\mu\gamma_\mu\Bigl(1-\gamma_5(\zeta_\parallel+
\vec \zeta_\perp\cdot\vec \gamma_\perp)\Bigr)
\]
where $\vec \zeta_\perp={2\over m}\langle \,\vec W -(\vec
W\cdot\vec n)\,\vec n\,\rangle$, and $\langle\,\cdot\,\rangle$
denotes an average. For helicity eigenstates $\vec
\zeta_\perp=0$ and $\zeta_\parallel$ equals twice the helicity so that
\begin{equation}
\rho={1\over 2} p^\mu\gamma_\mu(1\pm\gamma_5).\label{16}
\end{equation}
(\ref{16}) is identical to the expression for the density matrix
of the Dirac neutrino. However,
for
superpositions of different helicities $\vec \zeta_\perp\neq 0$
 which seems to suggest that even in the infinite
momentum limit some ``remains" of spin's components in directions
perpendicular to the momentum may be found.
Still,  there is no
contradiction with our analysis if we treat  $\vec {S}_p$
and not $\vec W$ as the relativistic spin operator. The
remains are those of $\vec W$ and not of $\vec {S}_p$. The
``transverse polarization" vector $\vec \zeta_\perp$ has to be
treated as a measure of superposition of the two helicites.

\section{LORENTZ CONTRACTION... OF WHAT?}
\label{what}

The decomposition of operators into even and odd parts can
be used for rewriting
the Dirac Hamiltonian in a form which is
 rather unusual but especially suitable for
investigation of its ultra-relativistic and massless limits. Let
us consider the angular velocity operator $\vec \omega$ defined
by (\ref{9}). For $m\neq 0$ and $\vec p\neq 0$ $\vec \omega$ does
not commute with $H$. Its even part, commuting with $H$,
 is given  (in ordinary units with $c\neq 1$) by
\[
\vec \Omega={c^2+m\,c^3\vec \gamma\cdot\vec n/|\vec p|\over
c^2+m^2c^4/\vec p\,^2}\,\vec \omega.
\]
We can see that $\vec \Omega$ reduces to $\vec \omega$ in both
limits. A Hamiltonian of a particle moving with velocity $\vec
v=c\vec \beta$ can now be expressed as
\begin{equation}
H=\Bigl(1+{m^2c^4\over c^2\vec p^2}\Bigr) \vec \Omega\cdot\vec
{S}_p
= \vec \beta^{-2}\vec \Omega\cdot\vec S
= \vec \beta^{-2}\vec \Omega\cdot\vec
{S}_p\label{NH}
\end{equation}
where each of the operators appearing in $H$ is even and
commuting with $H$. The limiting form $H=\vec \omega\cdot\vec S$
is characteristic of all massless fields, where for higher
spins the equation (\ref{9}) is still valid, but angular velocities
for a given momentum are smaller the greater the
helicity.

The new form of the Hamiltonian leads to the following interesting
observation \cite{ja}. Notice that for massless fields the
Hamiltonian can be written in either of the following two forms
\begin{equation}
H=\vec \omega\cdot\vec S\label{H1}
\end{equation}
or
\begin{equation}
H=\vec c\cdot\vec p=\vec v\cdot\vec p\label{H2}
\end{equation}
where $\vec v$ is the velocity operator for a general massless field
 ($c\vec \alpha$ in case of the
Dirac equation) and $\vec c=(\vec v\cdot\vec p)\vec p/\vec p\,^2$ is
its even part. We recognize here the classical mechanical rule for
a transition from a point-like description to the extended-object-like
one: linear momentum goes into angular momentum, linear velocity into
angular velocity, and vice versa. The third part of this rule (mass--moment
of inertia) can be naturally postulated as follows
\begin{equation}
H=m_k\vec c\,^2=I_k\vec \omega\,^2.\label{H3}
\end{equation}
where (\ref{H3}) defines the kinetic mass $(m_k)$ and the kinetic moment
of inertia $(I_k)$ of the massless field. The explicit form
of $I_k$ for massless fields of helicity $s=m-n$ [corresponding to
the $(m,n)$ spinor representation of $SL(2,C)$] is, in ordinary units,
\begin{equation}
I_k=\frac{s\hbar\vec p\cdot\vec S}{c \vec p\,^2}.\label{in}
\end{equation}
The  equation
\begin{equation}
I_k=m_k r_s^2
\end{equation}
characteristic, by the way, of circular strings (here with mass $m_k$)
defines some radius which is equal to
\begin{equation}
r_s=\frac{\hbar s}{|\vec p|}\label{rad}
\end{equation}
which can be expressed also as an (operator!) form of the
``uncertainty principle''
\begin{equation}
{|\vec p|} r_s={\hbar s}.\label{rad'}
\end{equation}
It is  remarkable  that this radius occurs also naturally   in the
 twistor formalism \cite{PR}. A twistor is a kind of a ``square root''
of generators of the Poincar\'e group on a light cone and belongs
to a carrier space of a representation of the conformal group.
It is known that although spin-0 twistors can be represented geometrically
by null straight lines, this does not hold for  spin-$s$, $s\neq 0$,
 twistors \cite{PR}. Instead of the straight line we get a congruence
of twisting, null, shear-free world lines, the so-called Robinson congruence.
A three-dimensional projection of this congruence consits of {\it circles\/},
whose radii are given exactly by our formula (\ref{rad}) (cf.~the footnote
at p.~62 in  \cite{PR}). The circles propagate with velocity of light in the
momentum direction and rotate in the right- or left-handed sense depending on
the sign of helicity.

The Robinson congruence picture is typical of {\it classical\/}
twistors.
It suggests that classical massless fields may be related naturally
to classical strings whose radii would have to
 be different for different inertial
observers. The quantized twistor formalism does not have such a
pictorial representation
since even for a spin-0 particle whose momentum {\it is given\/} no world
line exists, but additionally
because of  the difficulties
with the relativistic position operator.

The string-like picture of massless fields resulting from the
moment of inertia formulas and from their agreement with the
classical Robinson congruence is a physical  indication that a fundamental
role should be played in this context by the conformal group.
Indeed, a transition from one inertial reference frame to another not
only transforms the particle's four-momentum, but simultaneously
{\it rescales\/} the radius $r_s$ of the congruence.

Finally, the fact that the massive Dirac
Hamiltonian  written in terms of
even operators has the ``energy of precession'' form (\ref{NH}) suggests
that some kind of an
 extended structure can be associated also with massive
spinning particles. Structures of this type were constructed explicitly
by Barut and collaborators \cite{b1,b2,b3}.

\section{AN APPLICATION OF $\vec {S}_p$:
THE BELL THEOREM FOR DIRAC'S ELECTRONS}

Let us consider two electrons with opposite momenta
$\vec p_1$ and $\vec p_2=-\vec p_1$
(we choose,
in this way, a  center of mass reference frame).

The squared total even spin operator
\begin{equation}
(\vec {S}_{p_1}\otimes {1} +{1}\otimes\vec {S}_{p_2})^2\label{calk}
\end{equation}
in the helicity basis is given by the matrix
\[
\left(
\begin{array}{cccc}
(1+{m^2\over p_0^2}){1} & 0 & 0 & 0\\
0 & {m^2\over p_0^2}{1} & {m^2\over p_0^2}{1} & 0 \\
0 & {m^2\over p_0^2}{1} & {m^2\over p_0^2}{1} & 0 \\
0 & 0 & 0 & (1+{m^2\over p_0^2}){1}
\end{array}
\right).
\]
where {1} is the $4\times 4$ identity matrix and $p_0$ is
the energy of one of the particles.
Its eigenvalues are: $1+{m^2\over p_0^2}$, $2{m^2\over p_0^2}$
and 0. The first two correspond to the non-relativistic triplet
state and the third one to the singlet (degeneracies of the
eigenvalues are, respectively, 8, 4 and 4). An important property
of the definition (\ref{calk}) is the usage of {\it squared\/} even
operators (this is not the same as the even part of the ordinary squared
two-particle spin operator).

The singlet state
takes in the helicity basis $\Psi_\pm$
 the usual form
 \[
\Psi ={1\over \sqrt{2}}
(\Psi_+\otimes\Psi_-  -\Psi_-\otimes\Psi_+).
\]
In order to prove the Bell theorem we must calculate the singlet
state average
of an analog of the nonrelativistic operator
$
\vec a\cdot\vec \sigma\otimes\vec b\cdot\vec \sigma.
$
Here we find
\begin{equation}
\langle {\Psi}| \frac{\vec a\cdot\vec {S}_{p_{1}}}
{|s_{\vec a}|}\otimes \frac{\vec b\cdot\vec {S}_{p_{2}}}
{|s_{\vec b}|}
| {\Psi}\rangle=
-{1\over 4 |s_{\vec a} s_{\vec b}|}(\vec a_\|\cdot\vec b_\| +
{m^2\over p_0^2} \vec a_\perp\cdot\vec b_\perp ) \label{17}
\end{equation}
where the symbols $\|$ and $\perp$ denote projections on,
respectively, the momentum direction and the plane perpendicular
to it. For $\vec a$ and $\vec b$ perpendicular to $\vec p_1$
 (\ref{17}) equals $-\vec a\cdot\vec b$, the formula known from the
nonrelativistic quantum mechanics, and the Bell theorem can be
formulated. For other directions the formula (\ref{17}) differs from the
nonrelativistic one so might be used for an experimental verification
of the even spin concept \cite{Fry}.

\section{CONCLUSIONS}

We have shown that the algebra of a little group of any physical
four momentum is isomorphic to the algebra of the even spin
operator. This result explains qualitatively the fact that massless
fields can exist only in extremal helicity states since eigenvalues
of the even spin's components perpendicular to momentum tend to
zero in both infinite momentum and massless limits.
A physical origin of this phenomenon can be explained by the
Lorentz flattenning of the Dirac particle provided the particle
is extended. For massless fields (or ultra-relativistic electrons)
the flattenned picture can be naturally associated with the
classical Robinson congruence of null worldlines, leading to a
 string-like {\em classical limit\/} of spinning particles.
 In this way we have returned to the old problem of localization
of spinning particles \cite{3,DE,Pryce,Bacry} and have found another
argument for their extended structure and usage of noncommuting position
operators.

\end{document}